\documentstyle[sprocl]{article}

\input{psfig}

\def\Journal#1#2#3#4{{#1} {\bf #2}, #3 (#4)}

\def\NPB{{\em Nucl. Phys.} B}

\def\ZPC{{\em Z. Phys.} C}

\def\be{\begin{equation}}
\def\ee{\end{equation}}
\def\bea{\begin{eqnarray}}
\def\eea{\end{eqnarray}}

\begin{document}

\title{JET FRAGMENTATION AND MLLA}

\author{A. N. SAFONOV, for CDF Collaboration}

\address{Department of Physics, University of Florida,\\
Gainesville, FL 32611, USA\\
E-mail: safonov@phys.ufl.edu} 


\maketitle\abstracts{ Recent CDF results
on inclusive momentum distributions and multiplicities of particles in restricted cones around jets
are compared to predictions using the Modified Leading Log Approximation.
We found that MLLA gives a very reasonable description of jet fragmentation for a 
wide range of energies. Model parameters are extracted separately from the
multiplicity and from the shape of the momentum distributions and are found to agree. The ratio
of charged particle multiplicities in gluon and quark jets measured in the context of MLLA is compared
to the model-independent result and also found to agree.}

\section{Introduction and Theoretical Background}

Perturbative QCD calculations, carried out in the framework of the Modified Leading Log
Approximation~\cite{mlla} (MLLA), complemented with the Local Parton-Hadron Duality
Hypothesis~\cite{lphd}  (LPHD), predict the shape of the momentum
distribution, as well as the total inclusive multiplicity, of particles in jets. The MLLA is an
asymptotic calculation, which proves to be infrared stable, in the sense that the
model cutoff parameter $Q_{eff}$ can be safely pushed down to $\Lambda_{QCD}$. LPHD is responsible
for the hadronization stage and implies that hadronization is local and happens at the end of the
parton shower development. In its simplest interpretation, the model has one parameter
$K_{LPHD}$, the rate of parton-to-hadron conversion:
\begin{equation}
N_{hadrons} = K_{LPHD} \times N_{partons}.
\label{eq:murnf}
\end{equation}

The low cutoff parameter Qeff allows the inclusion of hadrons with low transverse momentum, which
constitute the majority of all hadrons in jets and could not be controlled in ordinary pQCD
with a conventional cutoff scale of the order of 1 GeV. In the most favorable scenario, $K_{LPHD}$
$\sim$ 1. If only charged particles are observed, one may expect $K^{charged}_{LPHD}$ to be between 
1/2 and 2/3.

In MLLA, momentum distributions and multiplicities in quark and gluon jets in a restricted cone of
size $\theta$ around the jet axis are functions of $E_{jet}\theta/Q_{eff}$~\cite{formulas} and differ 
by a factor $r$:
\begin{equation}
N^{q-jet}  ( \xi )  = \frac{1}{r} N^{g-jet}( \xi ), where {\  } \xi = \log{\frac{1}{x}},
x=p_{track}/E_{jet}
\label{eq:mu2}
\end{equation}

Jets at the Tevatron are a mixture of quark and gluon jets. Therefore,
\begin{figure}[t]
\centerline{\psfig{figure=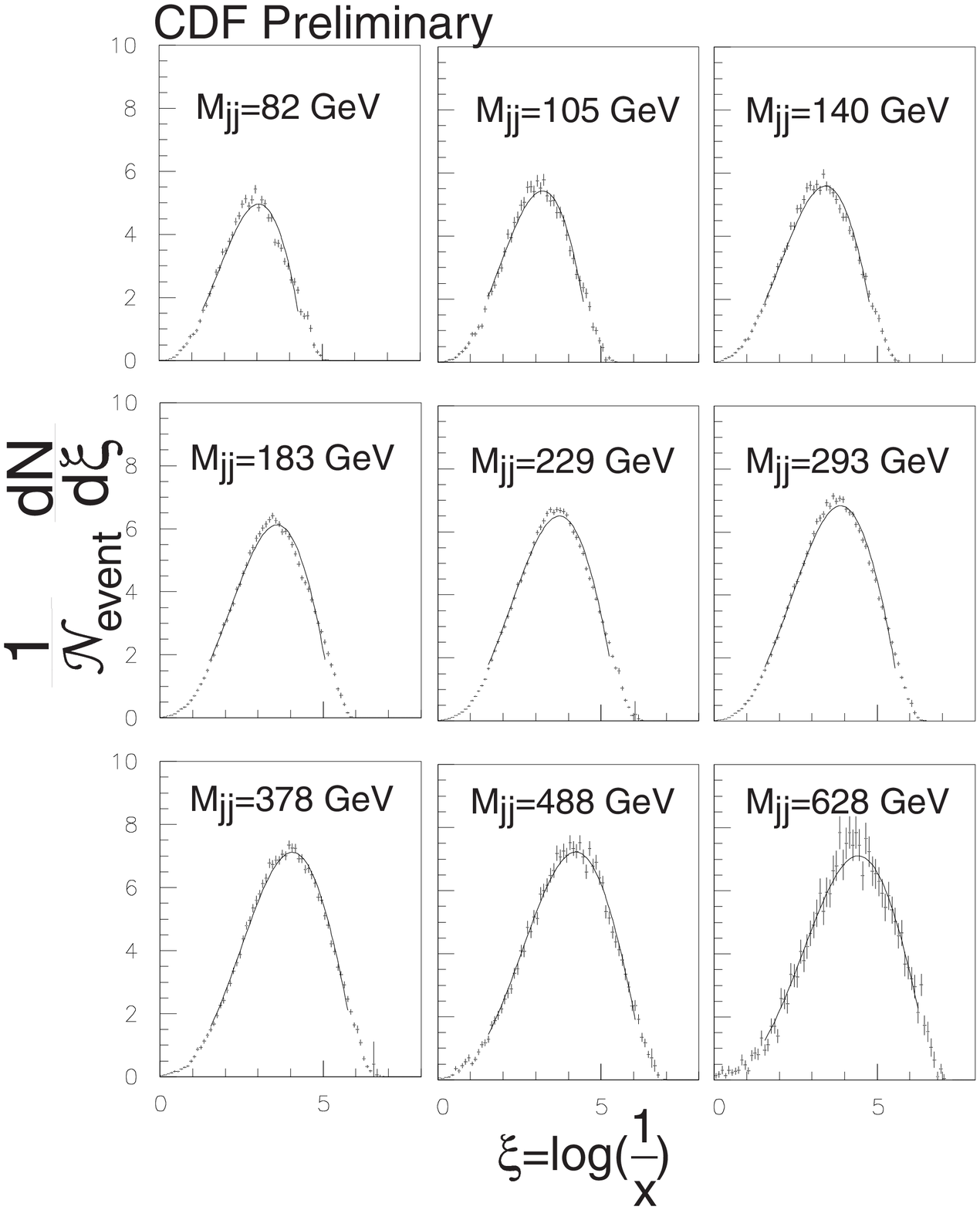,height=2.1in} 
\psfig{figure=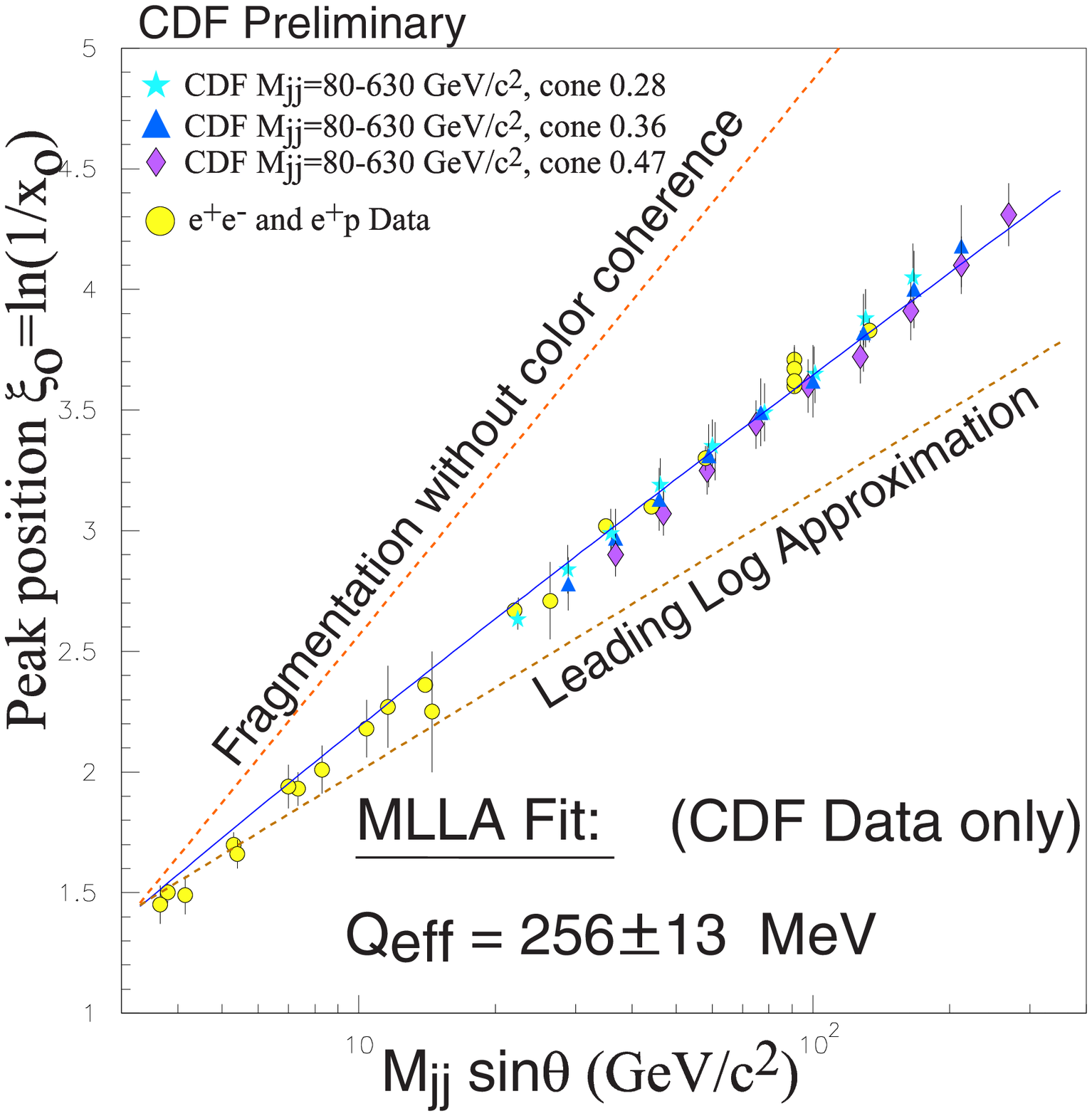,height=2.1in}}
\caption{Left: Momentum distribution of charged particles in dijet events. Fitted with MLLA function for
$Q_{eff}$ and $K$; Right: Peak position vs $M_{jj} \sin{\theta}$.  \label{fig:mom}}
\end{figure}
\begin{equation}
N^{charged}_{hadrons} ( \xi )  = K^{charged}_{LPHD} ( \epsilon_{g}+(1-\epsilon_{g})\frac{1}{r})
F^{nMLLA} N^{q-jet}_{part}( \xi ) = K  N^{q-jet}_{part}( \xi )  
\label{eq:mu3}
\end{equation}
where $\epsilon_g$ is the fraction of gluon jets in the events, the factor of $1/r$ reflects the difference
between gluon and quark jets, and, finally, factor $F^{nMLLA}$ accounts for the next-to MLLA
corrections to the gluon spectrum. Theoretical calculations~\cite{theor1} predict somewhat
different values of $F^{nMLLA}$, but all agree that $F^{nMLLA}$ has almost no dependence on the jet
energy in the region relevant to this analysis. We chose the average of the results above and used
the difference between predictions as a theoretical error: $F^{nMLLA}$=1.3$\pm$0.2. 
The same papers predict the value of $r$ to be between 1.5 and 1.8.

\section{Dijet Data Analysis}
We used data collected by CDF during the 1993-1995 running period. For the analysis, we selected 
events with two jets well balanced in transverse energy. Both jets were required to be in the 
central region of the detector to ensure reliable tracking reconstruction. To avoid biases, third 
and fourth jets were allowed if sufficiently soft. Data was further subdivided into 9 dijet mass 
bins (mea values were ranging from 80 to 630 GeV). Tracks were counted in restricted cones of sizes 0.28,0.36 and 0.47 around the jet axis.
Fig. 1(left) shows the inclusive mometum distribution of charged particles in jets for the 9 
dijet mass bins for the largest cone-size of 0.47. The data was fitted with Eq.(3) for 
the parameters 
$Q_{eff}$ and normalization $K$ (see Eq.(3)). The visual agreement is good; however, the $\chi^2$ is generally 
large indicating the importance of higher order and hadronization effects. 
We separately fitted the position of the peak of the momentum distribution for 27 combinations (9
dijet mass bins x 3 cone-sizes). In MLLA, the peak position depends on $Q_{eff}$ and is predicted
to have $E_{jet} \theta / Q_{eff}$ scaling. In Fig. 1(right), we plotted the dependence of the peak
position on $M_{jj} \sin{\theta}$ ($e^+e^-$ and $ep$ data is shown as well). 
Clearly, the predicted scaling is present in data.
Fig. 2(left) shows the 27 fitted values for $Q_{eff}$ as a function of the dijet mass. Taking into
account that the errors are dominated by correlated systematics, one can conclude that $Q_{eff}$ is not
absolutely universal, which again may be an indication of higher order effects. However, the scale
of the deviations is very moderate suggesting that these effects are not large. We report a value
for $Q_{eff}=240\pm40$ MeV (corresponding range shown as a band on Fig. 2(left)).
\begin{figure}[t]
\centerline{\psfig{figure=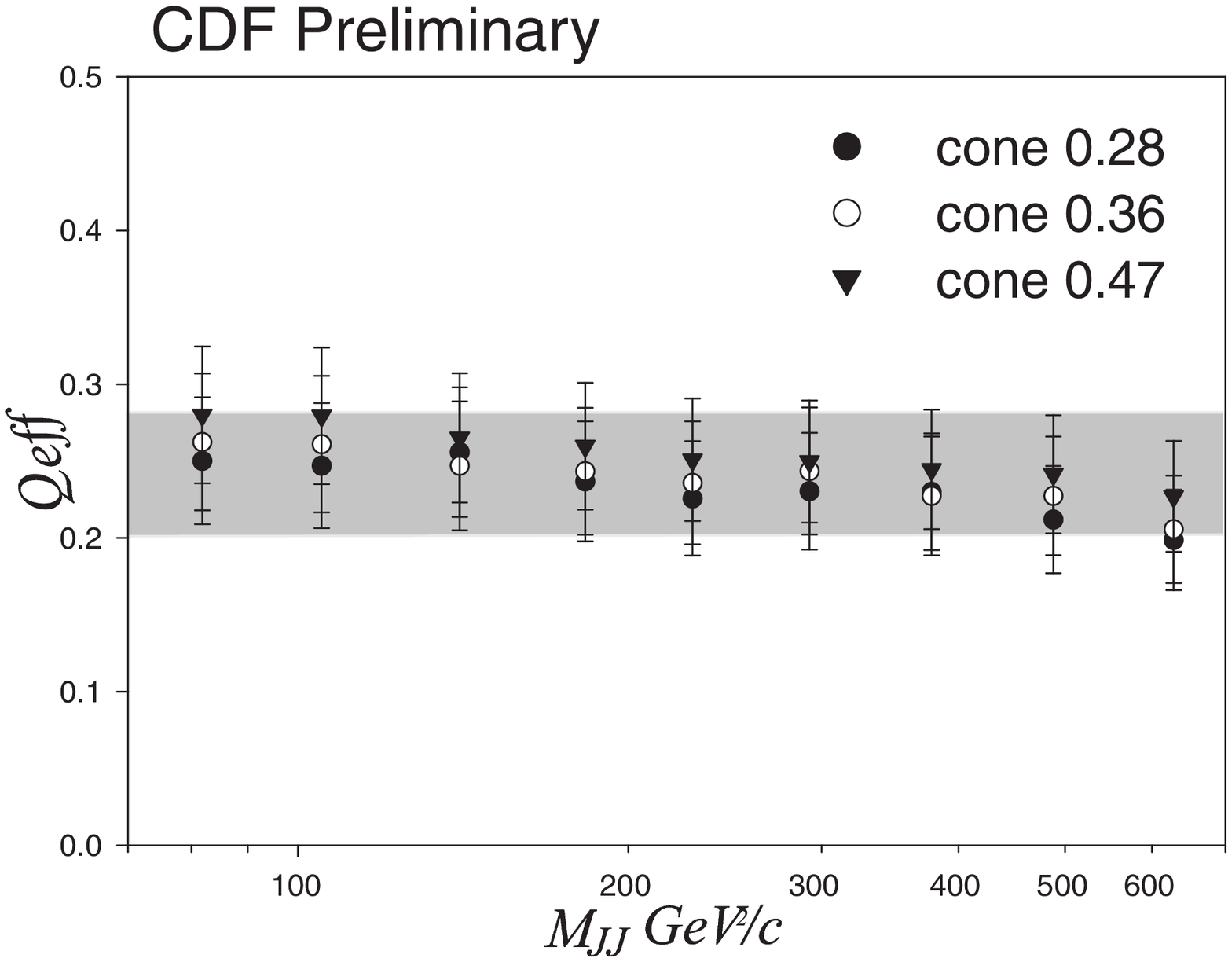,height=1.7in}
\psfig{figure=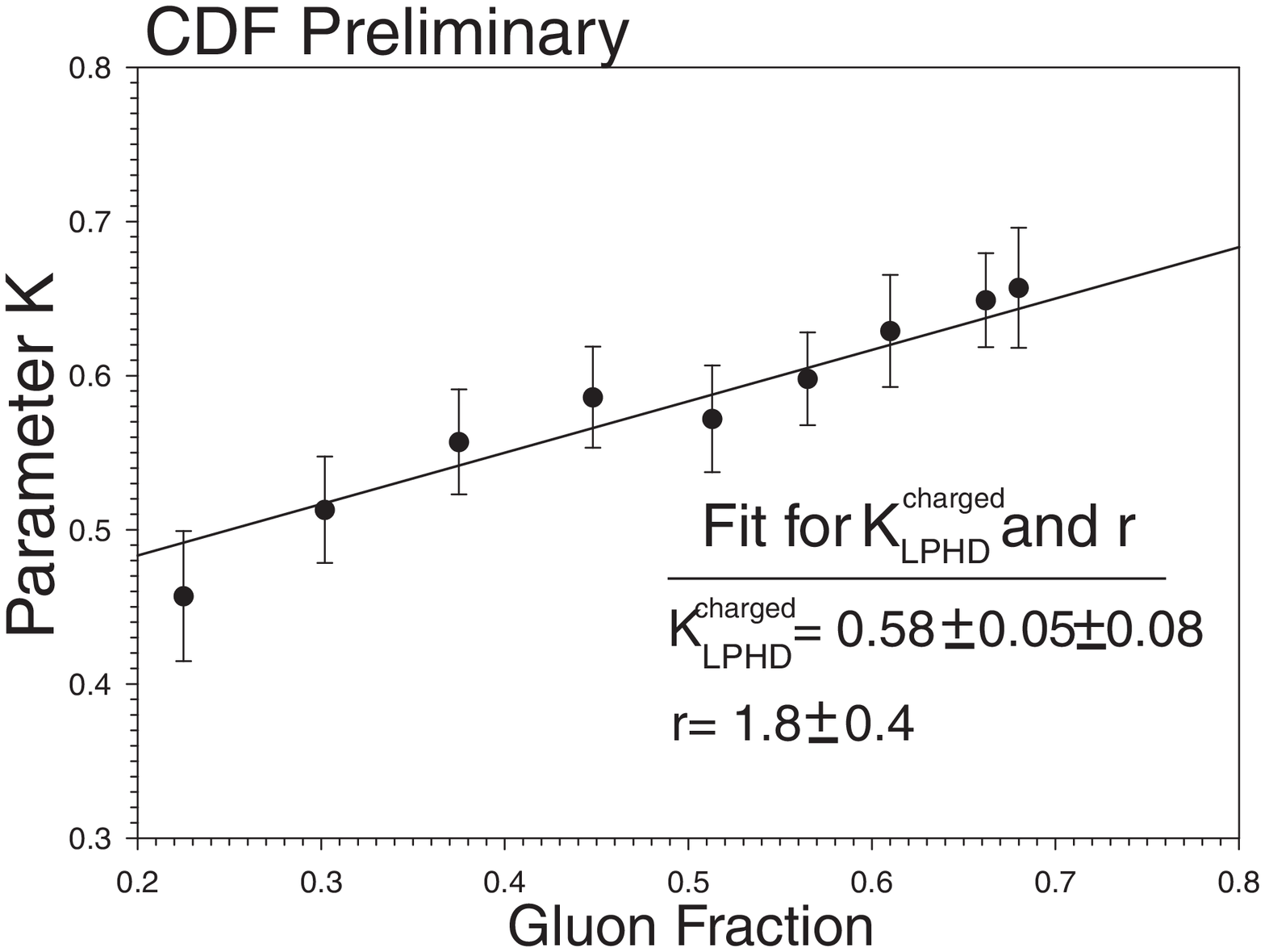,height=1.7in}}
\caption{Left: Fitted values of $Q_{eff}$  for 27 combinations (9 dijet mass bins x 3 cone-sizes). 
Errors are dominated by correlated systematic uncertainties; Right: Fit of the parameter $K$ 
for $K_{LPHD}$ and $r$. Cone 0.47. First error is combined statistical and systematic errors, 
the second one - theoretical error coming from $F^{nMLLA}$.    \label{fig:mom}}
\end{figure}
Analysis of the fitted parameter K allows an extraction of both $K_{LPHD}$ and r. 
According to (3), the dependence is linear.On Fig. 2(right), 
we plotted 9 values of K (corresponding to 9 dijet masses for the largest
cone-size 0.47) vs the gluon jet fraction (extracted using Herwig 5.6) in the 
events from respective dijet mass bins, as well as the results of the fit for $K_{LPHD}$ and $r$.
The same parameters can be extracted from the inclusive multiplicity using an integrated version of
Eq.(3). In this case, the extracted parameters will only rely on the total multiplicity and not on
the exact shape of the distribution. Fig. 3(left) shows the fit of data with MLLA predictions as well as
the fitted parameters $K_{LPHD}$ and $r$. It is remarkable that the two results are in such a good agreement.

\section{Model-independent Measurement of r}
We compared the multiplicity in dijet and $\gamma$-jet events (data selection was similar) to extract model-independent measurement of r. These samples have
very different fraction of gluon jets for the jet energies 40-60 GeV (roughly 60\% for dijets and
12\% for $\gamma$-jet, according to Herwig 5.6). The multiplicities measured for each of the samples and a
knowledge of the gluon jet fractions allowed us to extract $r$. Fig. 3(right) shows the measured r as a
function of the jet energy. We report $r$=1.75$\pm$0.11$\pm$0.15 in perfect agreement with MLLA result.
\begin{figure}[t]
\centerline{\psfig{figure=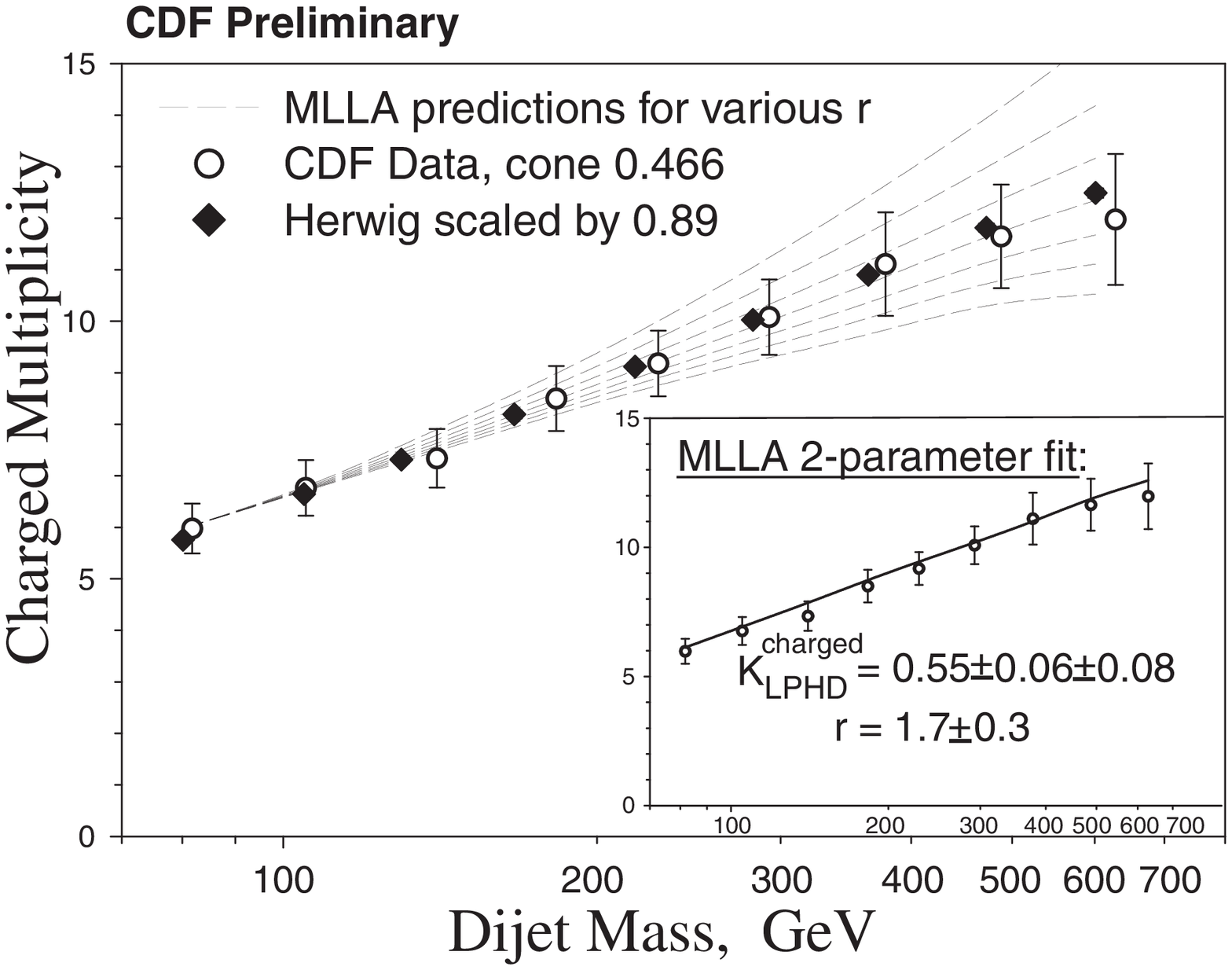,height=1.6in}
\psfig{figure=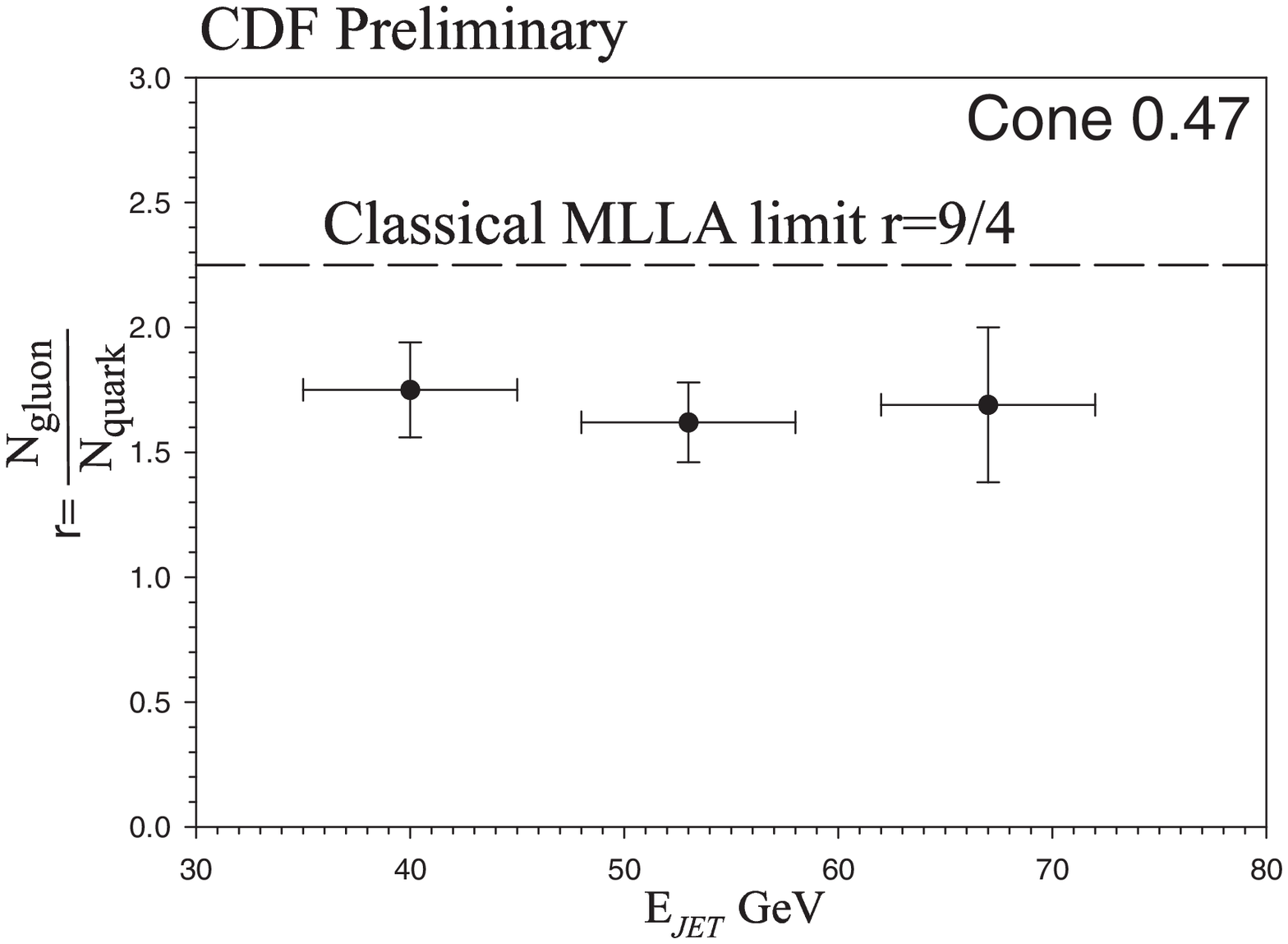,height=1.6in}}
\caption{Left: Charged particle multiplicity (per jet) as a function of the dijet mass. MLLA fit for
$K_{LPHD}$
and $r$; Right: ratio of charged multiplicities in gluon and quark jets based on comparison of the dijet and
$\gamma$-jet events.   \label{fig:mom}}
\end{figure}

\section*{Conclusion}
Results presented support the perturbative nature of jet fragmentation. The measured value of
$Q_{eff}$ allows a consistent description of the majority of particles. $K_{LPHD}$
and $r$ are not only self-consistence within the model, but also with model-independent result. 
$E_{jet}\sin{\theta}$ scaling is observed for the first time.

\end{document}